\documentclass[twocolumn]{aastex62}

\usepackage{graphicx}	
\usepackage{amsmath}	
\usepackage{amssymb}	
\usepackage{multirow}
\usepackage{newtxtext,newtxmath}

\usepackage[T1]{fontenc}
\usepackage{ae,aecompl}
\usepackage{comment}

\received{January 1, 2018}
\revised{January 7, 2018}
\accepted{\today}
\submitjournal{ApJ}

\shorttitle{VELA DL BNs}
\shortauthors{Huertas-Company et al.}

\begin{document}

\def \aj {AJ}
\def \mnras {MNRAS}
\def \pasp {PASP}
\def \apj {ApJ}
\def \apjs {ApJS}
\def \apjl {ApJL}
\def \aap {A\&A}
\def \nat {Nature}
\def \araa {ARAA}
\def \iaucirc {IAUC}
\def \aaps {A\&A Suppl.}
\def \qjras {QJRAS}
\def \na {New Astronomy}
\def \aapr {A\&ARv}
\def\lesssim{\mathrel{\hbox{\rlap{\hbox{\lower4pt\hbox{$\sim$}}}\hbox{$<$}}}}
\def\gtrsim{\mathrel{\hbox{\rlap{\hbox{\lower4pt\hbox{$\sim$}}}\hbox{$>$}}}}

\title{Deep Learning Identifies High-z Galaxies in a Central Blue Nugget Phase in a Characteristic Mass Range}

\correspondingauthor{Marc Huertas-Company}
\email{marc.huertas@obspm.fr}

\author{M. Huertas-Company}
\affil{Sorbonne Universit\'e, Observatoire de Paris, Universit\'e PSL, \\
CNRS, LERMA, \\
F-75014, Paris, France}
\affil{Sorbonne Paris Cit\'e, Universit\'e Paris Diderot, \\
F-75013, France}
\affil{Institut Universitaire de France}

\author{J. R. Primack}
\affiliation{Physics Department, University of California \\
Santa Cruz, CA 95064, USA}

\author{A. Dekel}
\affiliation{Physics Department, University of California \\
Santa Cruz, CA 95064, USA}
\affiliation{Racah Institute of Physics, The Hebrew University \\
Jerusalem 91904 Israel}

\author{D. C. Koo}
\affiliation{Department of Astronomy and Astrophysics, University of California \\
Santa Cruz, CA 95064, USA}

\author{S. Lapiner}
\affiliation{Racah Institute of Physics, The Hebrew University \\
Jerusalem 91904 Israel}

\author{D. Ceverino}
\affiliation{Institut fur Theoretische Astrophysik, Zentrum fur Astronomie, \\
Universitiat Heidelberg, Albert-Ueberle-Str. 2 \\
D-69120 Heidelberg, Germany}

\author{R. C. Simons}
\affiliation{Johns Hopkins University 3400 N Charles St. \\
Baltimore, MD, 21218, USA}

\author{G. F. Snyder}
\affiliation{Space Telescope Science Institute, 3700 San Martin Dr \\
Baltimore, MD 21218, USA}

\author{M. Bernardi}
\affiliation{Department of Physics and Astronomy, University of Pennsylvania \\
Philadelphia, PA 19104, USA}

\author{Z. Chen}
\affiliation{Shanghai Key Lab for Astrophysics, Shanghai Normal University \\
100 Guilin Road, 200234, Shanghai, China}

\author{H. Dom\'inguez-S\'anchez}
\affiliation{Department of Physics and Astronomy, University of Pennsylvania \\
Philadelphia, PA 19104, USA}

\author{C. T. Lee}
\affiliation{Physics Department, University of California \\
Santa Cruz, CA 95064, USA}
\affil{Sorbonne Universit\'e, Observatoire de Paris, Universit\'e PSL, \\
CNRS, LERMA, \\
F-75014, Paris, France}

\author{B. Margalef-Bentabol}
\affil{Sorbonne Universit\'e, Observatoire de Paris, Universit\'e PSL, \\
CNRS, LERMA, \\
F-75014, Paris, France}

\author{D. Tuccillo}
\affil{Sorbonne Universit\'e, Observatoire de Paris, Universit\'e PSL, \\
CNRS, LERMA, \\
F-75014, Paris, France}
\affil{MINES Paristech, PSL Research University \\
Centre for Mathematical Morphology, Fontainebleau, France}



\begin{abstract}

We use machine learning to identify in color images of high-redshift galaxies an astrophysical phenomenon predicted by cosmological simulations. This phenomenon, called the blue nugget (BN) phase, is the compact star-forming phase in the central regions of many growing galaxies that follows an earlier phase of gas compaction and  is followed by a central quenching phase.
We train a Convolutional Neural Network (CNN) with mock "observed" images of simulated galaxies at three phases of evolution: pre-BN, BN and post-BN,
and demonstrate that the CNN successfully retrieves the three phases in other simulated galaxies. We show that BNs are identified by the CNN within a time window of $\sim0.15$ Hubble times.
When the trained CNN is applied to observed galaxies from the CANDELS survey at $z=1-3$, it successfully identifies galaxies at the three phases. 
We find that the observed BNs are preferentially found in galaxies at a characteristic stellar mass range, $10^{9.2-10.3} M_\odot$ at all redshifts.
This is consistent with the characteristic galaxy mass for BNs as detected in the simulations, and is meaningful because it is revealed in the observations when the direct information concerning the total galaxy luminosity has been eliminated from the training set. This technique can be applied to the classification of other astrophysical phenomena for improved comparison of theory and observations in the era of large imaging surveys and cosmological simulations.

\end{abstract}

\keywords{galaxies: fundamental parameters, galaxies: high-redshift, galaxies: bulges}


\section{Introduction} \label{sec:intro}

Over the past years, we have acquired a detailed view of the statistical properties of galaxies at different cosmic epochs, thanks in particular to large scale imaging and spectroscopic surveys (e.g. SDSS;~\citealp{2000AJ....120.1579Y}, CANDELS;~\citealp{Koekemoer2011}). However, establishing causal connexions between galaxy populations remains an important challenge (e.g. \citealp{LillyCarollo2016}). This is obviously because of the timescales involved, which do not allow observations to follow the evolution of individual galaxies and also because of the degenerate link between commonly used observables and astrophysical processes.

This is particularly true for the processes leading to morphological transformations of galaxies, which remain largely unconstrained despite the large quantities of available data. A fundamental question, how bulges form and grow in galaxies at different cosmic times, is still largely debated. One of the reasons is that morphological observables extracted from images are rather simplistic and have essentially remained unchanged for many years. The characterization of galaxies is essentially limited to the prominence of the bulge and disk components based on the measurement of the central density (e.g. \citealp{2017ApJ...840...47B}), a parametric decomposition (e.g. \citealp{1968adga.book.....S, Peng2002}) or a ratio between enclosed light at different radii (e.g. \citealp{1996MNRAS.279L..47A}). The interpretation of these observables to constrain an assembly history is a very degenerate problem, i.e. there are many different processes that can lead to the same observables. \\

Our community is about to generate unprecedentedly large imaging datasets (e.g EUCLID, LSST, WFIRST). Hydro cosmological numerical simulations are also growing rapidly. It is thus timely to investigate alternative ways to extract a maximum amount of information from polychromatic images that might help break degeneracies with physics and improve the comparison between observations and simulated datasets. This is precisely the goal of this work. Ideally, one would like to have morphological measurements that directly correlate with some astrophysical process as predicted by theory and detected in simulations. That way, it would be possible to isolate objects from large surveys with high probability of experiencing a physical process and enable a more complete follow up. This is easily understandable for galaxy-galaxy mergers since it is a relatively well defined process associated with expected morphological features, at least at a first approximation. As a result, many efforts have been made to characterize merging galaxies from images (e.g. \citealp{2000ApJ...529..886C, 2008ApJ...672..177L}) and to calibrate their observability timescale to constrain the merger history (\citealp{2008MNRAS.391.1137L, 2017MNRAS.468..207S}). In that respect, it is important to calibrate with simulations that closely match the properties of the observed samples. For example, as shown in~\cite{2015ApJ...805..181C}, the morphological signatures of mergers at $z>1$ differ from those of mergers at $z\sim0$, and parametric classifications that robustly identify low-z mergers fail at $z>1$. 

Generalizing to other processes is less obvious since one needs to find the appropriate tracers from the multi wavelength pixel distribution. In recent years, there has been significant progress in the image processing community with the emergence of the so-called unsupervised feature learning techniques or deep learning (DL). These algorithms allow the user to automatically extract observables (or features) from the pixel space without any a-priori dimension reduction. As in many other disciplines, deep learning is rapidly being adopted in astronomy as well. It has been successfully used for several standard classification (e.g.\citealp{Huertas2015, 2015MNRAS.450.1441D,2017arXiv171105744D}) and regression problems (e.g~\citealp{2017arXiv171103108T}). We aim at investigating here an alternative way of using these advanced machine learning techniques to extract more physically relevant features from images and help establish a more solid link between theory and observations. \\

In this exploratory proof-of-concept work, we explore whether deep learning can be used to detect a phenomenon dubbed as blue nugget (BN), recently found in numerical simulations of high redshift galaxies.  Indeed these cosmological simulations \citep{2015MNRAS.450.2327Z, 2016MNRAS.458..242T, 2016MNRAS.457.2790T} reveal that a large fraction of the simulated galaxies undergo events of gaseous compaction, triggered, e.g., by mergers or counter-rotating inflowing streams, which leads to a central \emph{blue nugget} (BN) at a characteristic stellar mass of $10^{9.2-10.3} M\odot$. The BN phase in turn triggers a central gas depletion and central quenching of star formation, sometimes surrounded by an extended, freshly formed, gaseous, star-forming ring/disc. Most of the structural, kinematic and compositional galaxy properties undergo significant transitions as the galaxy evolves through the BN phase (\citealp{2015MNRAS.453..408C}, Dekel et al., in prep.). One way to investigate whether these gaseous compactions are frequent in the observed galaxies would be then to directly detect features in the data (stellar distribution in our case) unequivocally associated with the BN phase. This is what we attempt in this paper. One main advantage of high resolution numerical simulations over, for example, semi-analytical models or low-resolution large volume simulations, is that we can use them to generate realistic \emph{observed} simulated images for which the evolution history is known by construction (e.g. \citealp{2015MNRAS.451.4290S}). One can therefore isolate a sample of simulated galaxies in the BN phase, as well as in the pre-BN or post-BN phases. In this work, we use state-of-the art zoom-in cosmological simulations with high spatial resolution \citep{2014MNRAS.442.1545C} to generate mock images as observed by HST of galaxies in a BN phase. We then use this dataset to train deep neural nets and explore whether the network is able to automatically find morphological proxies associated with the different phases in the observed mock data. We then apply the trained network to observed CANDELS data. \\

The paper proceeds as follows. Sections~\ref{sec:sims} and~\ref{sec:data} describe the simulations and data used in this work. The main methodology is discussed in section~\ref{sec:train}. We show the main results on simulations and observations in sections~\ref{sec:res} and~\ref{sec:real} respectively.

\section{Simulations}
\label{sec:sims}
\subsection{Main properties of the simulations}

We use a set of zoom-in hydro cosmological simulations of 35 intermediate mass galaxies among which 31 are used in this work. The typical stellar mass of the simulated galaxies at $z\sim2$ is $10^{10}$ solar masses as shown in table~\ref{tbl:table1}. This is part of the VELA simulation suite which has been described and analyzed in several previous works \citep{2014MNRAS.442.1545C, 2015MNRAS.453..408C, 2015MNRAS.450.2327Z, Tacchella2015, 2016MNRAS.458.4477T, 2016MNRAS.458..242T}. We refer the reader to the aforementioned works for a detailed description of the simulations.  We summarize here only the most relevant properties. The initial conditions for the simulations are based on dark matter haloes that were drawn from dissipationless N-body simulations. The simulations were run with the AdaptiveRefinement Tree (ART) code \citep{1997ApJS..111...73K, 2003ApJ...590L...1K, 2009ApJ...695..292C}  and the maximum resolution is $17-35$ pc at all times, which is achieved at densities of $\sim 10^{4}-10^{3}$cm$^{-3}$. The code includes several physical processes relevant for galaxy formation: gas cooling by atomic hydrogen and helium, metal and molecular hydrogen cooling, photoionization heating by the UV background with partial self-shielding, star formation, stellar mass loss, metal enrichment of the ISM and stellar feedback. In particular, the high spatial resolution allows tracing the cosmological streams that feed galaxies at high redshift, including mergers and smooth flows, and they resolve the Violent Disk Instabilities (VDI) that governs high-z disc evolution and bulge formation~\citep{2009ApJ...703..785D}. This is important for this work focused on the growth of bulges and the reason why this small set of simulations is preferred to larger but lower resolution runs like Illustris. We recall that the gravitational softening for baryons in the Illustris series is of the order of $\sim 1kpc$ which means that any physical process that acts in smaller scales is unresolved. This is the case of the BN phase explored in this work.

However, as with all state-of-the art numerical simulations, the VELA simulations suffer from several limitations specially related to sub-grid physics. Like other simulations, the treatment of star formation and feedback processes still depends on uncertain recipes. The code assumes indeed a SFR efficiency per free fall time without following in detail the formation  of  molecules  and  the  effect  of  metallicity  on the SFR \citep{2012ApJ...753...16K}.  Additionally, no AGN feedback is yet included in the run used in this work. As a result, the full quenching observed in the data is not reached in many galaxies by the end of the simulations at $z\sim1$. Since we are more interested here in the BN phase that occurs when the galaxy is still star-forming, we do not expect that AGN will have a big impact on our results. However a color mismatch between simulated and observed galaxies might be expected. Besides that, as shown in \cite{2014MNRAS.442.1545C} and~\cite{2016MNRAS.458..242T}, the SFRs, gas fractions, and stellar-to-halo mass ratios are all close to the constraints imposed by observations, providing a better match to observations than earlier simulations.The uncertainties and any possible remaining mismatches by a factor of order 2 are comparable to the observational uncertainties. 

We stress that we are fully aware that the simulations present many limitations and that they are still very far from capturing all the complex physics of galaxy formation. This is mainly why the present work needs to be considered as a proof-of-concept work in that respect.  However, we are at a stage at which we can produce fairly realistic galaxies that capture some of the physical processes governing the assembly history, and we have good reasons to think that this will be improved in the future. This enables a comparison with observations in a more general way that we explore in this work. 

\subsection{Mock \emph{Candelized} images}
\label{sec:mock_images}
The full output of the simulation is saved at many time steps and analyzed at steps of $\Delta a=0.01$ of the expansion factor, which roughly correspond to $\sim 100$ Myrs at $z\sim2$. For every snapshot between $z\sim4$ and $z\sim1$ , we generate mock 2D images as they would be observed by the HST. They are generated using the radiative transfer code \textsc{sunrise}\footnote{sunrise is freely available at thttps://bitbucket.org/lutorm/sunrise.} \citep{2006MNRAS.372....2J,2010NewA...15..509J, 2010MNRAS.403...17J} by propagating the light of stars through the dust. We refer to \citealp{2015MNRAS.451.4290S} for details on the procedure followed.  

Very briefly, a spectral energy distribution (SED) is assigned to every star particle in the simulation based on its mass, age, and metallicity. The dust density is assumed to be directly proportional to the metal density predicted by the simulations. We set a dust-to-metals mass ratio of 0.4 (e.g. \citealp{1998ApJ...501..643D, 2002MNRAS.335..753J}), and the dust grain size distribution from updated by \cite{2007ApJ...657..810D}. \textsc{Sunrise} then performs dust radiative transfer using a Monte Carlo ray-tracing technique.  As each multiwavelength ray emitted by every star particle and HII region (according to its SED) propagates through the ISM and encounters dust mass, its energy is probabilistically absorbed or scattered until it exits the grid or enters one of the viewing apertures (\emph{cameras}). The output of this process is then the SED at each pixel in all cameras. For this run we set 19 cameras from which five are fixed with respect to the angular momentum vector of each galaxy, seven are fixed in the simulations coordinates and the remaining seven are fully random at each time step. The camera numbers are summarized in table~\ref{tbl:table2}.

Finally, from these data cubes, we create raw mock images by integrating the SED in each pixel over the spectral response functions of the CANDELS WFC3 filters ($F160W$, $F125W$ and $F105W$) in the observer frame.   Images are then convolved  with the corresponding HST PSF at a given wavelength. We finally add a random real noise stamp taken from the CANDELS data. This ensures that the galaxies are simulated at the same depth than the real CANDELS data and also that the correlated noise from the HST pipeline is well reproduced. We call this process \emph{CANDELization}. 

For each 3D snapshot ($\Delta a=0.01$), we therefore generate 19 different 2D images corresponding to the 19 different camera orientations. The resulting dataset corresponds to approximately $\sim10000$  images in each of three filters. Even if the CANDELS filters probe the optical rest-frame up to $z\sim3$, we included galaxies up to $z\sim4$ since the most intense compaction events tend to happen at higher redshift in the VELA simulations. Given that the gas fractions (stellar to halo mass relations) are slightly underestimated (overestimated) in the simulations as previously stated, including higher redshift is justified and increases the size of our training set. We have checked however that the main results of the paper remain unaltered if only galaxies up to $z\sim3$ are used.  We emphasize that the same procedure has been used to generate mock JWST galaxies in the different filters and therefore a similar analysis as the one presented in this work can be applied to this dataset in order to prepare JWST observations.

\begin{table}
     \begin{center}
     \begin{tabular}{|l|l|}
     \hline
    
        camera number & orientation \\
     \hline
cam00/02 & Angular momentum face-on (opposite directions)\\ 
cam01/03 & Angular momentum edge-on (opposite directions) \\
cam04 & Angular momentum 45 degrees \\
cam05/06/07 & Fixed to x,y and z axis simulation box\\
cam08-11 & Random (same simulation coord. for all snapshots) \\
cam12-18 & Fully random \\
     \hline
     \hline
     \end{tabular}
     \end{center}
     \caption{Explanation of the 19 camera orientations used to generate mock 2D images from the simulations.}
     \label{tbl:table2}
 \end{table}
 
 \section{Data}
\label{sec:data}
We also use HST observational data to test our model in section~\ref{sec:real}. We use CANDELS images in the three infrared filters (F105W, F125W, F160W) from the 2 GOODS fields (North and South, \citealp{Grogin2011,Koekemoer2011}). The selection is based on the morphological catalog presented in \cite{Huertas2015}, which is essentially a selection of the brightest galaxies (F150W<24.5) from the official CANDELS catalogs (\citealp{Guo2013}, Barro et al. 2017). This is required to have enough S/N to measure morphological information from images. For this work, we select only galaxies in the redshift range $1-3$ to match the simulated redshift range.  As shown in \cite{Huertas2016}, the sample is mass complete down to $10^{9}$ solar masses at $z\sim1$ and $10^{10}$ at $z\sim3$. We restrict our analysis to galaxies more massive than $10^9$ solar masses to have enough statistics and match the typical stellar masses from the simulations. The sample might therefore suffer from incompleteness at high redshift. This is not critical for the illustrative purpose of this work.    

In addition to the reduced images, we also use official CANDELS redshifts \citep{2013ApJ...775...93D} which are a combination of photometric redshifts computed with several codes and spectroscopic when available. Stellar masses and star-formation rates from SED fitting are also used. Stellar masses are computed through SED fitting using the best redshift adopting a \cite{2003PASP..115..763C} IMF. SFRs are computed by combining IR and UV rest-frame luminosities (\citealp{1998ApJ...498..541K,2005ApJ...625...23B}) with \cite{2003PASP..115..763C} IMF (see \citealp{2011ApJS..193...13B} for more details). The following relation was used: $SFR_{UV+IR}=1.09\times10^{-10}(L_{IR}+3.3L_{2800})$. Total IR luminosities are obtained using \cite{2001ApJ...556..562C} templates fitting MIPS $24\mu m$ fluxes and applying a \emph{Herschel based} recalibration. For galaxies undetected in $24\mu m$, SFRs are estimated using rest-frame UV luminosities \citep{2011ApJ...742...96W}. We also compute for the selected dataset the central mass density ($\Sigma_1$) following the methodology of \cite{2017ApJ...840...47B}. 

\section{Methods: Training the network}
\label{sec:train}

\subsection{Training set: using the simulation metadata to label images}
\label{sec:labels}
The final goal is to train a deep neural network to identify, from the mock images, the BN phase (and consequently the pre and post-BN phases as well).  As put forward by previous analysis of the same simulated dataset~\citep{2015MNRAS.450.2327Z}, almost all the simulated galaxies seem to evolve in three characteristic phases. They go from diffuse to compact star-forming objects through wet gas compaction to then quench in the central regions and build a central bulge that will in most of the cases rebuild a surrounding stellar disk. We notice that the intensity of the compaction depends on stellar mass, and while most of the simulations go through a BN phase, only the most massive become compact star-forming galaxies. 

\begin{figure*}
\begin{center}
$\begin{array}{c c}
\includegraphics[width=0.45\textwidth]{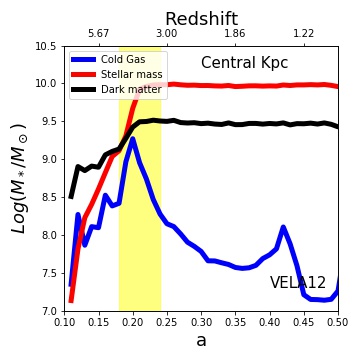} & \includegraphics[width=0.45\textwidth]{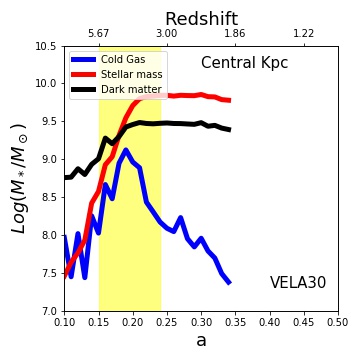} \\ 
 \includegraphics[width=0.45\textwidth]{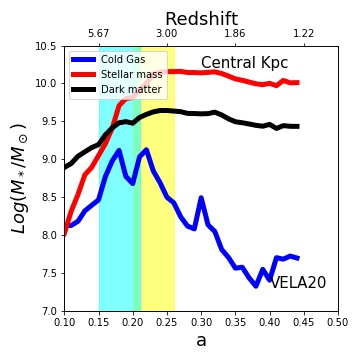}  &  \includegraphics[width=0.45\textwidth]{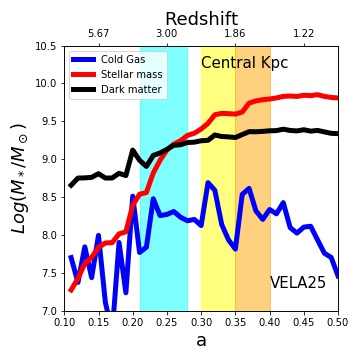}\\
\end{array}$
\caption{Definition of the different phases. Both the cold gas and the stellar mass in the central kpc are used to define the BN phase. The blue and red line show the evolution of the cold gas and stellar masses in the central kpc as a function of time. The black line is the dark matter mass. (Adapted from Zolotov et al. 2015). The yellow shaded region shows the main BN event as defined in this work (see text for details). The second and third  events are shown in cyan and orange respectively. The ranking refers to the intensity of the event and no to the time of occurrence (see text for details). Each panel shows a different galaxy. The top panels show clear examples of massive galaxies with one unique BN phase. The bottom panels show more complex cases with more than one BN event. } \label{fig:comp_def}
\end{center}
\end{figure*}

As part of the training set, we then first define these 3 phases for all the galaxies in the simulation. The identification of the 3 phases is performed in an individual basis for each galaxy using the gas density evolution in the central galactic regions as explained in \cite{2015MNRAS.450.2327Z} and Dekel et al. (in prep). Basically we identify the \emph{peak of the BN phase} as the time at which the gas density in the central kpc is maximum. We define the end of the BN phase when the central stellar density stops increasing, which is a signature that star-formation has been quenched in the center of the galaxy. The onset of the BN phase is considered to start when the central gas density starts to increase toward the BN peak. Naturally, this is more complicated than selecting the peak. In our current approach the selection is done by eye using also the 2D projection of the gas density to confirm the choice. Figure~\ref{fig:comp_def} shows the cold gas and stellar mass evolution in the central kpc for some galaxies for illustrative purposes. We also show the dark matter content in the central kpc. The key take-away message from these plots is that compaction is not always well defined and that it comes in many different flavors. There are for instance some \emph{clean} cases as VELA12 in which there is a single peak of the gas mass. However, there are other cases such as VELA25 for which the peak is not so pronounced and identifying the boundaries of the BN phase is not obvious and somewhat arbitrary. Notice also that many galaxies experience several BN phases as for example discussed in \cite{2016MNRAS.457.2790T}. In this work, we define a maximum of 3 BN phases for each galaxy as shown in table~\ref{tbl:table1}. Table~\ref{tbl:table1} summarizes indeed the redshifts of the BN phase of all galaxies analyzed. This is to say that the network that will be trained needs to somehow capture this heterogeneity in the process. It is important to keep this in mind when analyzing the results. \\
As can be seen in table~\ref{tbl:table1}, in the simulations, the BN phase tends to happen at a characteristic galaxy stellar mass $\sim 10^{9.2-10.3}M_\odot$ (e.g. \citealp{2015MNRAS.450.2327Z}). Given the existing correlation between mass and luminosity, this implies that there is a brightness gradient between pre-BN, BN and post-BN, with pre-BN galaxies being generally fainter than post-BN. The difference in luminosity also implies a difference in S/N when the HST noise is added. A deep learning approach, as the one used in this work, has the unique power to automatically extract the optimal tracers from the data to minimize the classification error. It also implies a risk since the network can potentially use any available information. In our case, given the properties of the training set, there is therefore a potential risk that the network uses the S/N difference existing between the different phases to classify. Since we do not want the network to learn based on S/N but rather learn characteristic features of the BN phase, we artificially shuffle the magnitudes of the galaxies given to the network. To do so, before adding the noise (see section~\ref{sec:mock_images}), we associate a random magnitude to all snapshots in the $F160W$ filter ($19-25$ in order to match the CANDELS distribution). This way, galaxies in the different phases have similar luminosities and S/N distributions. By doing so, the characteristic mass information is also washed out preventing the network from using that information to learn. We will discuss the effect of this choice in section~\ref{sec:real}. We remark that all other properties are kept unchanged. It includes obviously the spatial distribution of pixels which measure the degree of compactness and also the relative luminosities in each filter which is correlated with the SFR.

We thus use this 3-class classification (pre-BN, BN or post-BN)  to associate a unique label to every simulated image. Pre-BN includes all galaxies before experiencing any compaction event i.e. with a redshift larger than the maximum of ($z^1_{onset}, z^3_{onset}, z^3_{onset}$). Galaxies in the BN phase are the ones with redshift between $z^y_{onset}$ and $z^y_{post}$, with $y={1,2,3}$. Finally, all remaining images are labelled as post-BN. So galaxies with several compaction events are classified as \emph{post-BN} between two events. As a result of this labelling process, every mock image has an associated label corresponding to its evolutionary phase. The final dataset consists therefore of $\sim 10000$ labelled images with the simulation assembly history that will be used to train and test a convolutional neural network model.

\begin{figure*}
\includegraphics[width=0.95\textwidth]{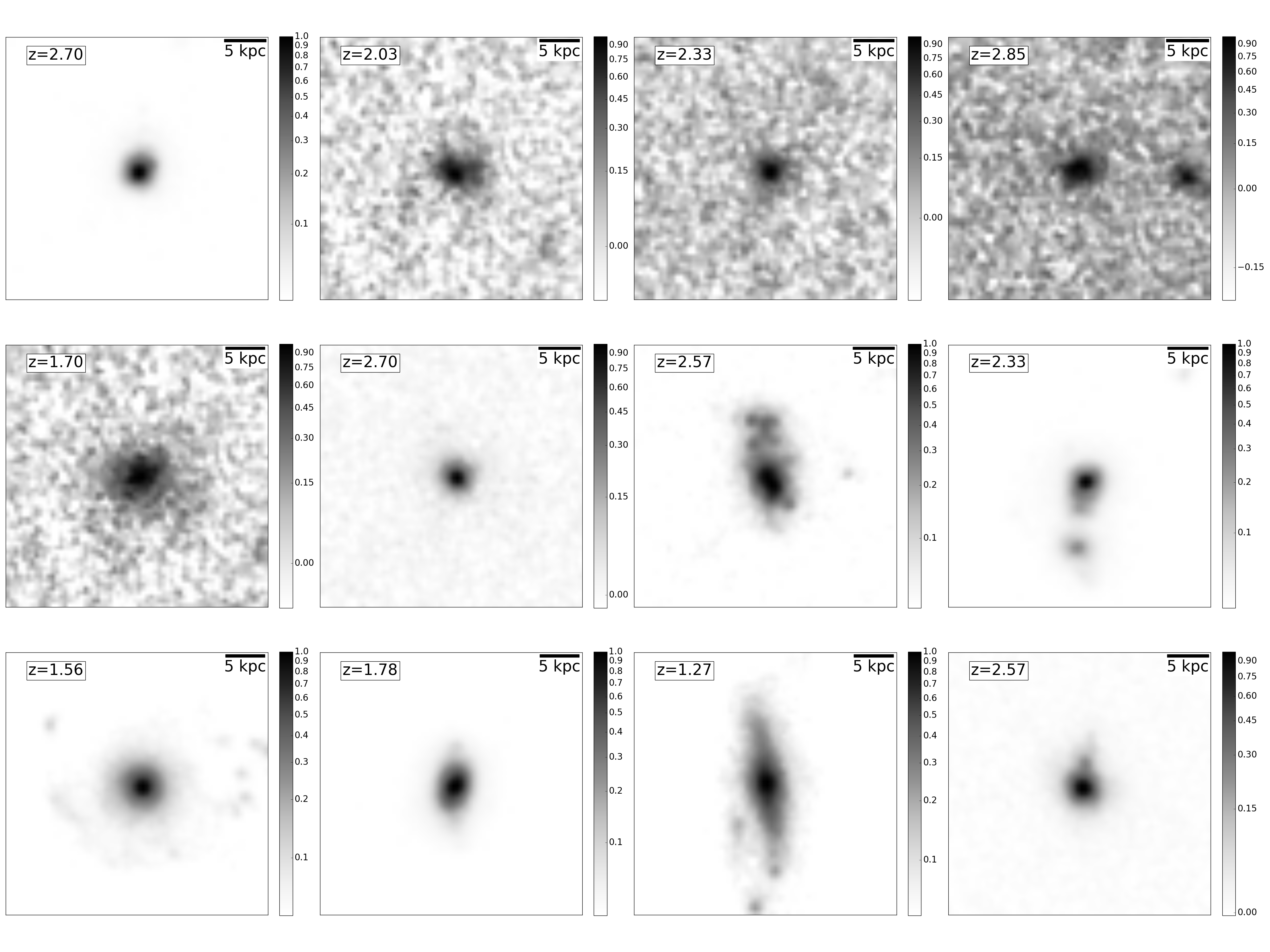}  \\ 
\caption{Random examples of simulated $F160W$ \emph{Candelized} images in the 3 phases discussed in this work. The image size is $3.8^{"}\times3.8^{"}$. The top row shows pre-BN galaxies, the middle row are galaxies in the BN phase, and the bottom row are post-BN objects. The images have been rescaled so that they span the same range of luminosities in the 3 phases. } \label{fig:ex_stamps}
\end{figure*}

Figure~\ref{fig:ex_stamps} shows some random example stamps of galaxies in the three phases in the HST/WFC3 F160W filter. Pre-BN galaxies generally look smaller and post-BN tend to have a diffuse disk structure. However, no obvious visual difference is apparent. This underlines the challenge of this work, which is to train a CNN capable of distinguishing between the different phases. 

\begin{table*}
     \begin{center}
     \begin{tabular}{|c|c|c|c|c|c|c|c|c|}
     \hline
    
        simulation & $z^1_{onset} $ & $z^1_{post}$ &  $z^2_{onset} $ & $z^2_{post}$ & $z^3_{onset} $ & $z^3_{post}$ & $Log M_*/M_\odot$ & $Log M_*/M_\odot$\\
         &  & &    &  &  & & $z=z_{comp}$  & $z=2$\\
     \hline
 VELA01& 1.86& 1.38& --& --& --& --& 10.05& 9.39\\ 
VELA02& 1.70& 1.00& --& --& --& --& 9.72& 9.32\\ 
VELA03& 3.00& 1.94& 1.27& 0.96& --& --& 9.47& 9.70\\ 
VELA04& 2.23& 1.63& 1.50& 1.17& --& --& 9.18& 9.07\\ 
VELA05& 1.38& 1.08& --& --& --& --& 9.47& 9.09\\ 
VELA06& 5.25& 3.17& 2.57& 1.86& --& --& 9.60& 10.42\\ 
VELA07& 3.55& 2.57& 4.88& 3.35& --& --& 10.39& 10.83\\ 
VELA08& 2.23& 1.50& 0.96& 0.69& --& --& 9.79& 9.79\\ 
VELA09& 4.00& 3.00& 1.63& 1.33& --& --& 9.73& 10.09\\ 
VELA10& 3.17& 2.13& 1.44& 1.13& --& --& 9.59& 9.83\\ 
VELA11& 4.00& 2.85& 2.12& 1.70& --& --& 9.67& 10.05\\ 
VELA12& 4.56& 3.17& --& --& --& --& 9.98& 10.33\\ 
VELA13& 2.85& 2.03& --& --& --& --& 9.76& 10.06\\ 
VELA14& 2.33& 1.56& --& --& --& --& 10.26& 10.19\\ 
VELA15& 2.70& 2.13& 1.70& 1.38& --& --& 9.70& 9.77\\ 
VELA17& 7.33& 3.55& 3.76& 2.57& --& --& 9.63& --\\ 
VELA19& 9.00& 4.56& 2.70& 2.13& --& --& 9.75& --\\ 
VELA20& 4.00& 2.85& 5.67& 3.76& --& --& 10.33& 10.62\\ 
VELA21& 3.55& 2.57& 4.88& 3.35& 7.33& 4.56& 10.51& 10.65\\ 
VELA22& 4.88& 3.55& --& --& --& --& 10.02& 10.67\\ 
VELA25& 2.33& 1.86& 3.76& 2.57& 1.86& 1.50& 9.89& 9.91\\ 
VELA26& 3.17& 2.13& 5.25& 3.55& --& --& 9.82& 10.25\\ 
VELA27& 2.23& 1.70& 3.35& 2.57& --& --& 9.90& 10.01\\ 
VELA28& 1.63& 1.22& --& --& --& --& 9.71& 9.51\\ 
VELA30& 5.67& 3.17& --& --& --& --& 9.87& 10.25\\ 
VELA32& 7.33& 4.00& --& --& --& --& 9.71& 10.52\\ 
VELA33& 4.88& 3.00& 3.35& 2.45& 2.33& 1.78& 9.61& 10.73\\ 
VELA34& 3.00& 1.78& 4.26& 2.70& --& --& 10.06& 10.32\\ 
     \hline
     \hline
     \end{tabular}
     \end{center}
     \caption{Summary of the BN phases for all simulated galaxies used in this work. For each galaxy we show the redshift(s) at which the BN phase(s) were identified to start ($z_{onset}$) and end ($z_{post}$). We also indicate the stellar mass of the galaxy when the main BN phase occurs as well as the stellar mass at $z=2$. A dash (-) means that the simulation did not run until $z\sim2$.}
     \label{tbl:table1}
 \end{table*}

\subsection{Architecture}
\begin{figure*}
\begin{center}
\includegraphics[width=0.95\textwidth]{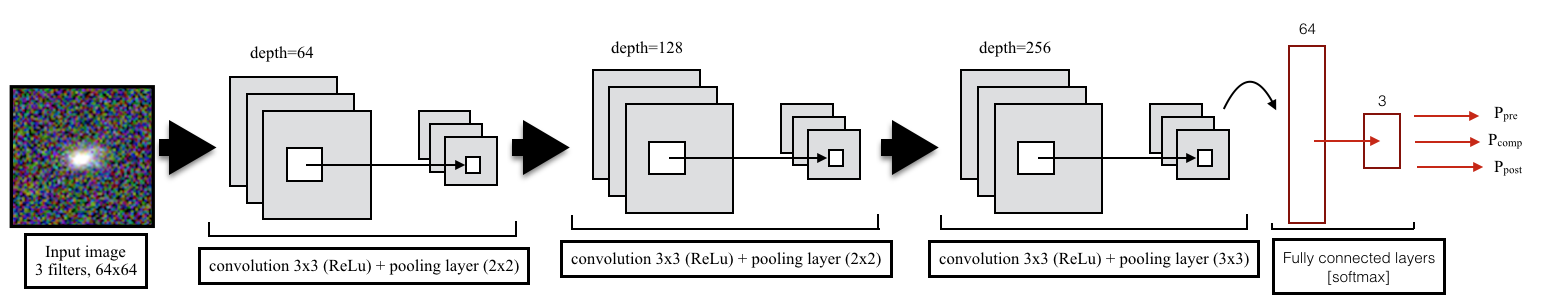} \\
\caption{Architecture of the deep network used for classification in this work. The network is a standard and simple CNN configuration made of 3 convolutional layers followed by pooling and dropout.  } \label{fig:arch}
\end{center}
\end{figure*}

We use a very simple sequential CNN architecture with only 3 convolutional layers followed by 2 fully connected layers implemented in Keras\footnote{https://keras.io/} with a Theano backend (figure~\ref{fig:arch}). The main reason to use a relatively shallow network is the limited size of the training set. The architecture is inspired by previous configurations which were successful in detecting strong lenses in space-based images \citep{2018arXiv180203609M} and also for classical morphological classification~\citep{2017arXiv171105744D}. We then add 2 fully connected layers to perform the classification. The last layer has a \emph{softmax} activation function to ensure that the 3 outputs behave like probabilities and add to one. We use a \emph{categorical crossentropy} as loss function and the model is optimized with the \emph{adam} optimizer.   

The network is fed with images (fits format) of fixed size  ($64\times64$ pixels) with 3 channels corresponding with the 3 main NIR CANDELS filters (F160W, F125W and F105W). We also tried to include bluer filters ($F850LP$), but the results do not change significantly.  For simplicity in this illustrative work, we thus used the 3 redder filters since the pixel scale is the same and hence no interpolation is required. In principle the number of filters could be increased without any significant modification of the methodology. The input size is a trade-off between properly probing the galaxy outskirts ($\sim30$ kpc in the redshift range $1-3$) and having a small enough number of input parameters to prevent overfitting. 

In addition to this, we also use standard techniques to avoid overfitting at first order. Firstly after each convolutional layer we apply a 50\% dropout. Secondly, we include a Gaussian noise layer at the entrance of the network to avoid that the model learns from the noise pattern given that our training set is small. Finally, we use real-time data augmentation. Images are randomly rotated (within 45 degrees), flipped and slightly off centered by 5 pixels maximum at every iteration so that the network does never see exactly the same image.

 \subsection{Training and validation strategy}

One obvious limitation we face in this work is that our training dataset is made of only $\sim28$ galaxies. Even though we increase the number of available images by using different camera orientations as well as data augmentation, there is a potential risk that the network learns how to identify the different phases for this particular set of galaxies without generalizing. To avoid this situation, we have designed a custom training strategy which slightly differs from the classical approaches typically used in machine learning. 

Among the 28 galaxies, we use 24 galaxies for training (i.e. $\sim9000$ images), 2 for real-time validation during the training and 2 additional completely independent galaxies for testing at the end of the training process. It is important to keep in mind that, when we say 2 galaxies it does not mean 2 images. Each galaxy corresponds to the full assembly history of the galaxy between $z=4$ and $z=1$ with 19 images at each time step. Therefore the test and validation sets contain $\sim1000$ galaxies each. 

\begin{figure}
\begin{center}
\includegraphics[width=0.45\textwidth]{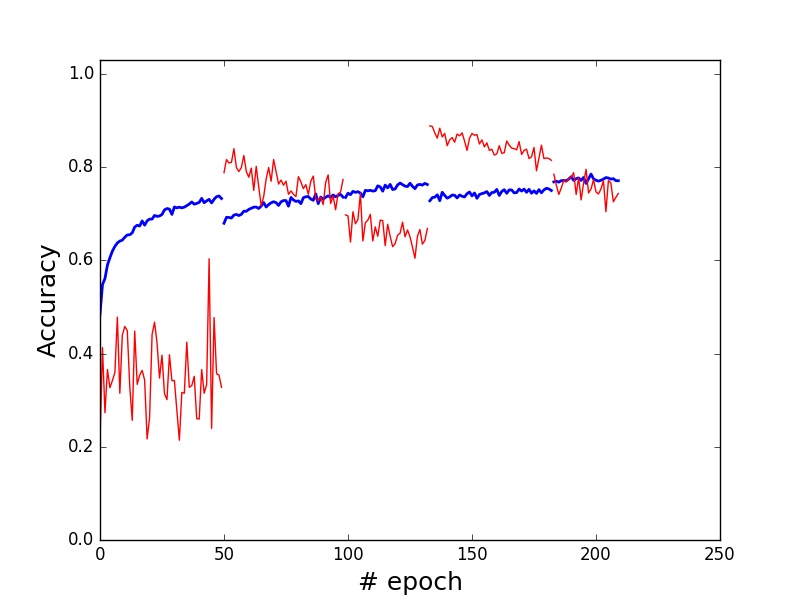} \\
\caption{Learning history resulting from the strategy described in the text. The blue solid line shows the accuracy on the training set and the red solid line is the accuracy for the validation set. Every 50 epochs the validation and training datasets are modified which explains the discontinuities. The accuracy on the validation is generally unstable because it is only made of 2 galaxies. See text for details. } \label{fig:learn_histo}
\end{center}
\end{figure}

We then train for a maximum of 250 epochs. The novelty is that every 50 epochs we move 2 galaxies from the training set to the validation sample and add the validation galaxies to the training. This helps the network not to overfit on the first sample of 24 galaxies while training for enough number of epochs to enable convergence. The 2 test galaxies are obviously never used in the process. Finally, in order to have more than 2 galaxies to test the classification accuracy, we repeat 5 times the training procedure just  described, using two different galaxies for the test sample at every run. The final test dataset thus contains 10 galaxies, classified with 5 slightly different models. Figure~\ref{fig:learn_histo} illustrates the learning history parametrized by the evolution of the accuracy as a function of the number of epochs of one of the five runs for illustration purposes. We plot the accuracy computed on the training and validation datasets. As expected the training curve monolithically increases and reaches roughly an accuracy of $80\%$. Notice however some small discontinuities every 50 epochs corresponding with the modification of the training set. The fact that the discontinuity is small suggests that small modifications of the training sample do not significantly alter the network performance. In other words, there is no over-fitting. The validation curve shows a particular behavior. This again is consequence of the adopted training strategy. Every 50 epochs there is a clearly noticeable jump. The break is larger than for the training because the validation is only made of 2 galaxies and the sample is fully changed every 50 epochs. So the break somehow reflects the accuracy variation between galaxies which can go from 100\% for some galaxies to $\sim 60\%$. As previously stated, compaction is not a very well defined process and some galaxies have complex assembly histories with multiple BN phases. The red curve also presents large jumps between epochs. This is also most probably a consequence of the size and redundancy of the sample. Given that there are 19 images per snapshot a change in the classification of a few snapshots implies big changes in the accuracy value.

\section{Results} 
\label{sec:res}

In this section we analyze the classification accuracy. We use for that purpose the test dataset (10 galaxies) which was not used in the training process (see section~\ref{sec:train}) throughout all the section.

\subsection{Detection of BNs}

\begin{figure}
\begin{center}
$\begin{array}{c}
\includegraphics[width=0.45\textwidth]{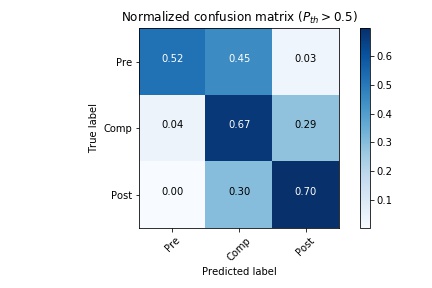}  \\
\includegraphics[width=0.45\textwidth]{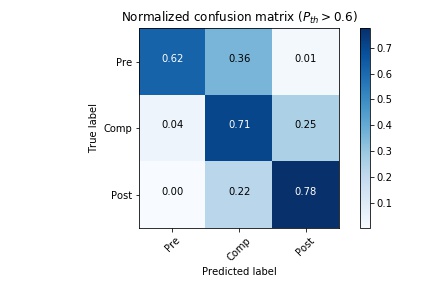}  \\
\includegraphics[width=0.45\textwidth]{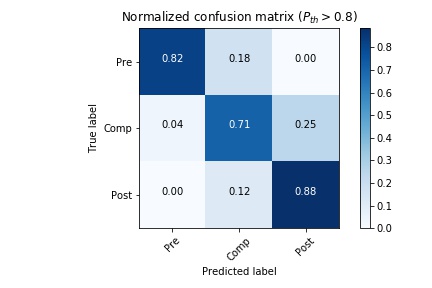} \\

\end{array}$
\caption{Normalized confusion matrix of the 3-label classification on a test dataset not used for training nor validation. The y-axis shows the true class from the simulation metadata, the x-axis is the predicted class. From top to bottom, we show the effect of increasing the probability threshold to select the galaxies belonging to a given class. } 
\label{fig:conf_matrix}
\end{center}
\end{figure}

We first analyze the average accuracy of the trained model to detect pre-BNs, post-BNs and BNs. The global accuracy, defined as the fraction of images correctly classified, computed on the test dataset is $\sim70\%$, which means that $30\%$ of the objects are misclassified. This is certainly not very high. Recall however that there is a significant amount of redundancy in the test set. It is helpful to look into more detail to better understand what is going on before drawing conclusions. We first compute a standard confusion matrix showing the relation between input and output classes (figure~\ref{fig:conf_matrix}) for different probability thresholds.  At the lower probability threshold (0.5) most of the confusion comes from true pre-BN (or post-BN) that are predicted as BN. This is probably because, as previously stated, the compaction event is not very well defined. The duration and intensity strongly depend on the galaxy. As expected, the degree of contamination decreases when a higher probability threshold is used to select galaxies. At the highest threshold (0.8) 25\% of true BNs are predicted to be post-BN. In fact one should keep in mind that the test set contains snapshots in steps of $\Delta_a=0.01$. A galaxy might be mis-classified as post-BN just before the compaction event ends for example or where there are multiple compactions closely followed in time, reducing the accuracy of the classification. However the classification might still be meaningful. 

\begin{figure*}
\begin{center}
$\begin{array}{c c}
\includegraphics[width=0.41\textwidth]{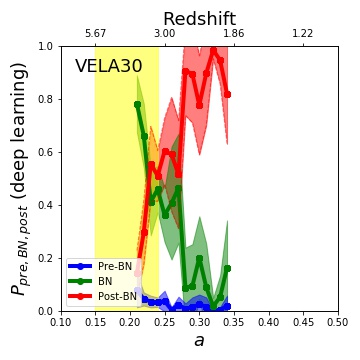} & \includegraphics[width=0.41\textwidth]{plots/comp_V30.jpg} \\
\includegraphics[width=0.41\textwidth]{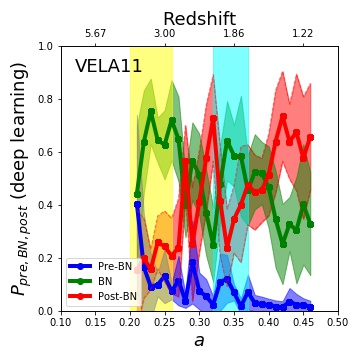} & \includegraphics[width=0.41\textwidth]{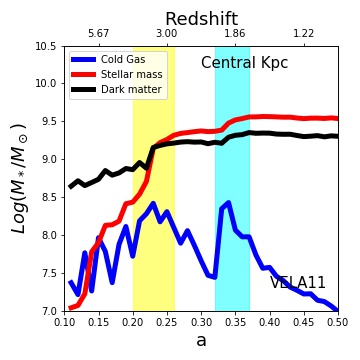} \\
\includegraphics[width=0.41\textwidth]{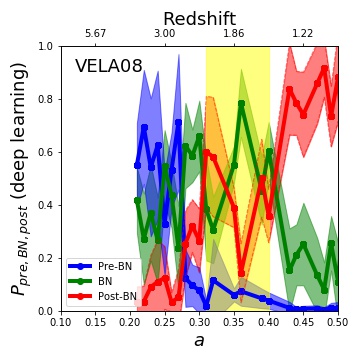} & \includegraphics[width=0.41\textwidth]{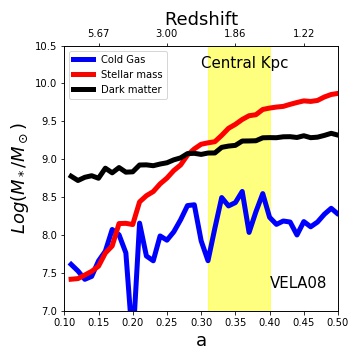} \\
\end{array}$
\caption{Examples of predictions on a test sample of increasing complexity. The left column shows the mean probability of being in pre-BN (blue solid line), BN (green solid line) and post-BN (red solid line) predicted by the CNN.The shaded regions around the lines indicate the scatter due to different camera orientations. The right column shows the input simulation metadata used to define the phases as in Figure 1. The yellow and cyan shaded regions show the primary and secondary BN phases.} \label{fig:pred_time}
\end{center}
\end{figure*}

To investigate this further, in figure~\ref{fig:pred_time} we plot the output probabilities for a subset of individual galaxies from the test sample as a function of time. In this figure, the lines show the average probability over all camera orientations at a given snapshot. The shaded regions show the scatter due to different camera orientations. For illustration purposes, we show three cases with increasing complexity. The first example (VELA30) has one single intense BN phase. VELA11 is less massive and has 2 events of smaller intensity. Finally VELA08 is a low-mass galaxy with a very weak compaction.These three examples bracket the diversity of assembly histories the network needs to capture.  As can be seen, there is a good correlation between the evolution of the probability values and the evolutionary phase. We observe that typically the probability of pre-BN tends to decrease before the compaction event, while the compaction probability increases. Towards the end of the BN phase, the probability of post-BN increases. This is true even for galaxies with complex assembly histories. This result indicates two main things. Firstly, it shows that the machine has learned somehow that there is a sequential order between the 3 phases. This is not obvious since all images were randomly included in the training process with random luminosities and, as seen in table~\ref{tbl:table1}, the BN phase can happen at very different redshifts and can have very different durations. Secondly, it shows that despite the relatively low global accuracy, the confusion seems to come essentially from the snapshots taken at the transition phases. This is important because it means that when the machine misclassifies it is not fully random. The misclassification therefore is a reflection of the difficulty to define the different phases. It is also worth noticing that the scatter due to different camera orientations is generally not large ($\sim0.1-0.2$ in terms of probability). It suggests a mild impact of the projection in the classification accuracy.

\subsection{Impact of camera orientation}

\begin{figure}
\begin{center}
$\begin{array}{c}
\includegraphics[width=0.45\textwidth]{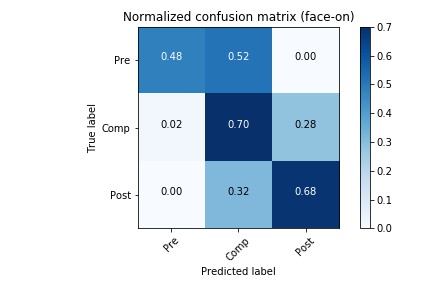}  \\
\includegraphics[width=0.45\textwidth]{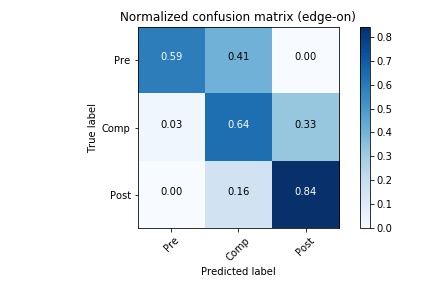}  \\
\includegraphics[width=0.45\textwidth]{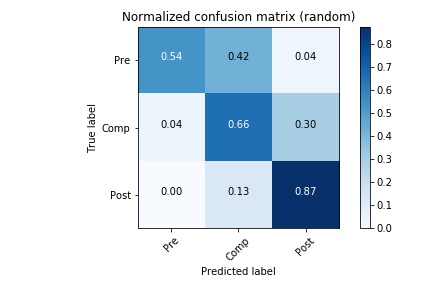} \\

\end{array}$
\caption{Same as previous figure but the confusion matrix is shown for different camera orientations. Top: Face-on (cam00/02), Middle: Edge-on (cam01/03), Bottom: Random (cam13+).} 
\label{fig:conf_matrix_cam}
\end{center}
\end{figure}

\begin{figure}
\begin{center}

\includegraphics[width=0.45\textwidth]{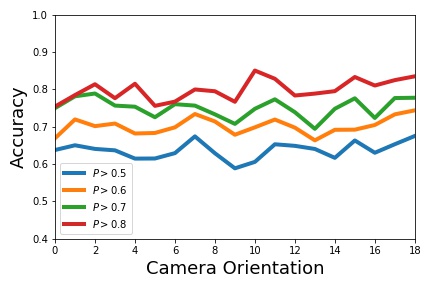}  \\

\caption{Measured accuracy on the test dataset as a function of the camera orientation. The numbers indicate the orientation (see Table 2). The different colors indicate different probability thresholds as labeled. The accuracy does not depend on the camera orientation. } 
\label{fig:acc_camera}
\end{center}
\end{figure}

We investigate this further in figure~\ref{fig:conf_matrix_cam}, which shows the confusion matrix divided by camera orientation. Despite some statistical fluctuations, no significant differences are appreciated as already suggested by the results shown in figure~\ref{fig:pred_time}. This is also quantified in figure~\ref{fig:acc_camera}, which shows the global accuracy as a function of the camera number (see table~\ref{tbl:table2} for an explanation of the different numbers). The figure confirms that there is no systematic trend with the orientation. The global accuracy increases equally for all cameras when the probability threshold is increased.

\subsection{Calibration of observability timescales}
\label{sec:obs_time}

\begin{figure}
\begin{center}
 \includegraphics[width=0.45\textwidth]{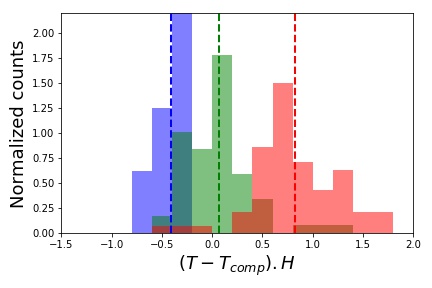} \\
\caption{Observability of the BN phase with the calibrated classifier. The histograms show the distributions of time (relative to the Hubble time at the time of the peak of the BN phase). The blue, green and red histograms show the pre-BN, BN and post-BN phases. The dashed vertical lines show the average values for each class with the same color code. Despite some overlap, the classifier is able to establish temporal constraints on the different phases. The darker regions indicate overlapping histograms. } \label{fig:time_comp}
\end{center}
\end{figure}

In fact, in view of applying the model to real data, probably the most interesting property to investigate is whether we can calibrate the observability timescales of the features learned by the classifier. In other words, what is the typical time window in which the network detects BNs. This is important because it allows us to better interpret the classification in terms of an evolutionary sequence and also to compute a \emph{BN rate} from the observations as usually done for mergers typically.  To do so, we take the test sample and classify all galaxies in the 3 classes according to the output probabilities. We simply add each image to the class of maximum probability and require that the probability value is larger than 0.5. We then compute, for each galaxy, the time difference with the main BN phase (computed as a fraction of the Hubble time at the BN peak, i.e $1/H(t)$, $H(t)$ being the Hubble constant. Figure~\ref{fig:time_comp} shows the histograms for the 3 classes. We confirm that the 3 classes tend to probe a different regime although with some overlap as expected from the results of the previous sections. Pre-BN galaxies are on average selected $\sim0.40/H(t)$ before the event and post-BN galaxies are typically observed $\sim0.80/H(t)$ after the compaction. The galaxies classified are centered on the BN phase ($0.05\pm0.3$ Hubble times).

Although there is some overlap between the different histograms, it is worth noticing that all galaxies which passed the BN phase by more than half a Hubble time are classified as post-BN galaxies. Also there are no galaxies classified as BN or pre-BN objects for which the event is more than $\sim0.5$ Hubble times away. This means that our classifier can indeed establish some temporal constraints regarding the BN phase based only on the stellar distributions.  This is extremely important because it shows that there is a temporal sequence implied in the classification. So when applied to real data one can more easily interpret the results in terms of evolution as will be discussed in section~\ref{sec:real}.

\subsection{Inside the network}

\begin{figure*}
\begin{center}
\includegraphics[width=0.90\textwidth]{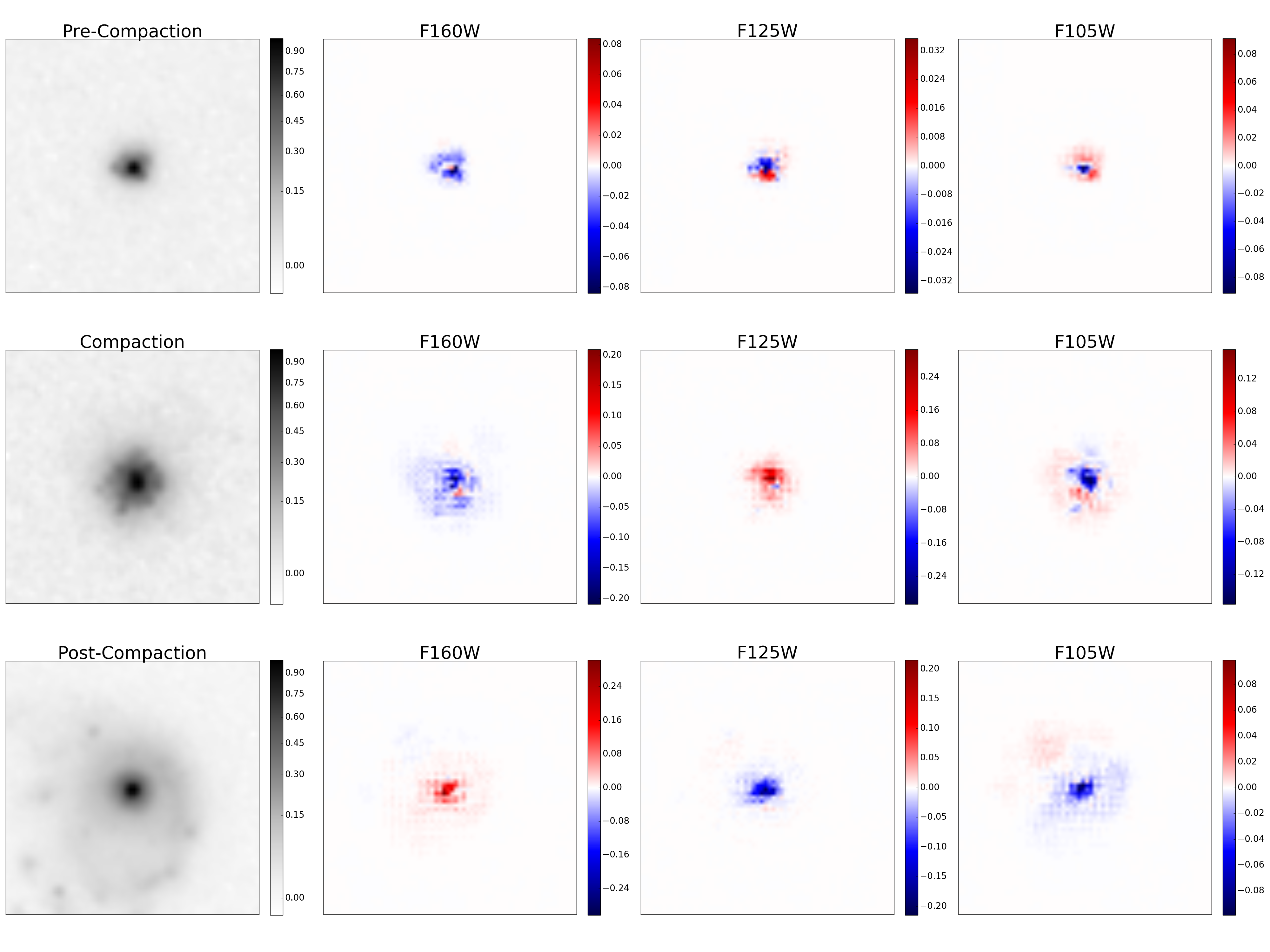} \\
\caption{Integrated gradients output of the model. Each row shows a galaxy in a different stage (Pre-compaction, Compaction, Post-Compaction). The left column is the original image and the second, third and fourth columns show the integrated gradients for the different filters. The network automatically detects the pixels belonging to the galaxy and used all of them to make the decisions.} \label{fig:int_grad}
\end{center}
\end{figure*}

An important caveat of the machine learning approach presented above is that it somehow behaves as a black box. It is thus difficult to precisely determine what are the features the machine is using to decide the output classification. This is a general problem for all deep learning applications. However, there exist more and more \emph{network interrogation} techniques which identify the pixels in the input image that most contributed to the final classification. One recent method is called integrated gradients~\citep{2017arXiv170301365S}. It is based on the measurement of the differences between gradients computed by the network in an input image as compared to a test image (usually a blank image with only zeros). We tested this method in our model and computed the integrated gradients for some of the galaxies. Figure~\ref{fig:int_grad} shows one example for each class. The interpretation is not straightforward. However some useful information can be extracted from this exercise. It is interesting to see that the model automatically segments all the pixels belonging to the galaxy and takes the decision based on all the galaxy pixels. It also means it understood there is no information in the noise and confirms that the model is not over-fitting on the noise pattern. 
Also, as pointed out in previous works, after the BN phase a gaseous disk is often built in the simulations \citep{2015MNRAS.450.2327Z,2016MNRAS.458..242T}. The bottom panels of the figure show clearly that the machine detects the diffuse disk component even if faint and probably uses this information to make the decision concerning the post-BN and sometimes the BN phase.  For galaxies in the BN phase, the relevant pixels are more concentrated in the center since the galaxies are generally more compact  as the obvious signature of this phase. It is also worth noticing that the gradient tends to have values of different sign in the center and in the outskirts as if the machine was using difference in flux between the center and the outskirts to classify. This is expected since the compaction event is by definition accompanied by a burst of central star-formation and the sSFR profiles evolve from decreasing to rising, indicating quenching outside-in in the pre-BN phase and inside-out in the post-BN phase \citep{2016MNRAS.458..242T}. The model is capturing all these correlations automatically. This is the strength of the presented methodology. Although the information that can be extracted from integrated gradients is quite limited at this stage, it is reasonable to think that interrogation techniques will become more advanced, and therefore there is potentially information that can be learned from a post-processing of the model outputs in the future.

\section{Identifying blue nuggets in the observations}
\label{sec:real}

We now apply the model to the HST/CANDELS sample presented in section~\ref{sec:data}. We simply cut stamps around the selected galaxies in the three infrared filters ($F160W$, $F125W$, $F105W$) and classify them into three classes using the trained models. Since 10 models were produced (see section~\ref{sec:train}), we use each of them to classify all galaxies. Each real galaxy has therefore 10 different classifications using slightly different models. We then compute the average probability to increase the robustness of the classification. We stress that there is a general good agreement between the different models which confirms that the classification does not strongly depend  on the specific subset of simulated galaxies used for training. The typical scatter in the probability values is of the order of $\sim0.1$. \\

\begin{figure*}
\includegraphics[width=0.95\textwidth]{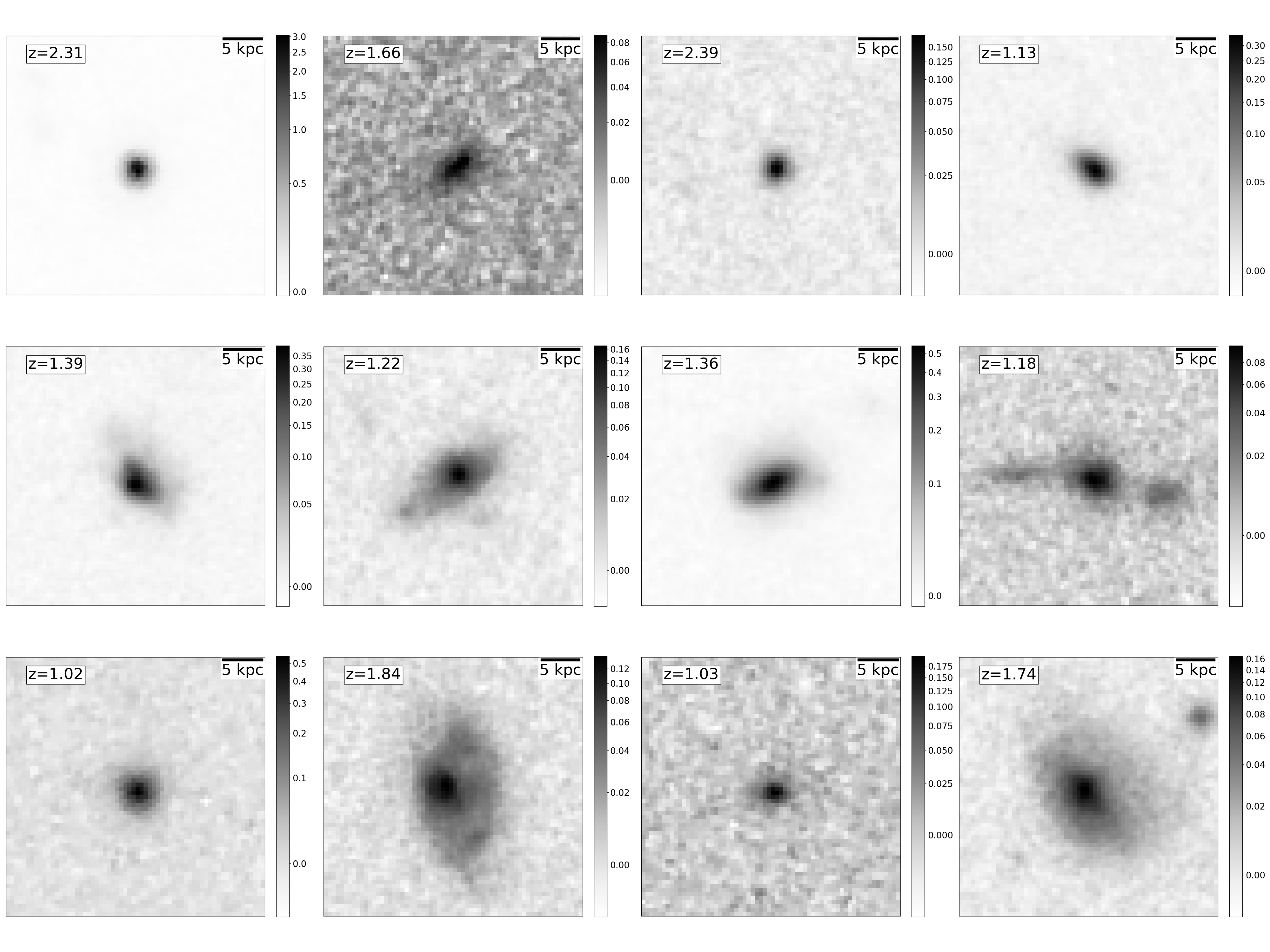}  \\ 
\caption{Random examples of $F160W$ CANDELS images in the 3 phases discussed in this work. The image size is $3.8^{"}\times3.8^{"}$. The top row shows pre-BN galaxies, the middle row are galaxies in the BN phase, and the bottom row are post-BN objects. } \label{fig:ex_stamps_CANDELS}
\end{figure*}

\begin{figure*}
\begin{center}
$\begin{array}{c c c}
 \includegraphics[width=0.33\textwidth]{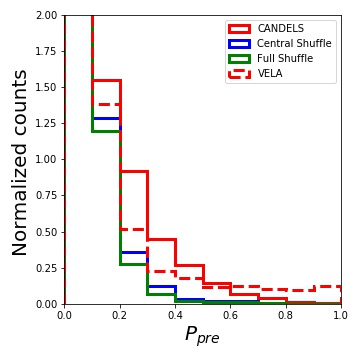} & \includegraphics[width=0.33\textwidth]{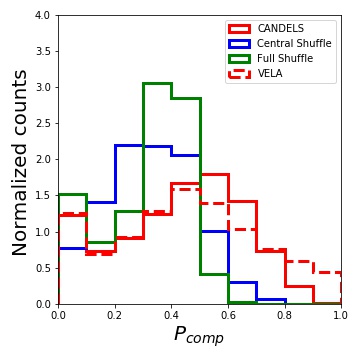} & \includegraphics[width=0.33\textwidth]{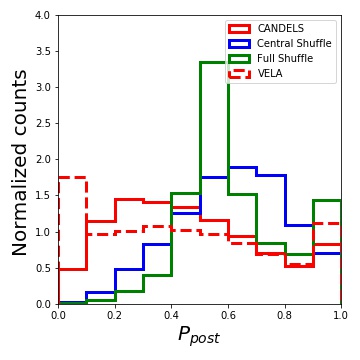}\\
\end{array}$
\caption{ Impact of shuffling the pixels on the output probability distributions. From left to right we show the pre-compaction, compaction and post-compaction probability distributions. The red solid lines show the distribution for the original CANDELS images. The blue (green) lines show the same distributions when the central (outskirts+central) pixels are shuffled. For reference, we also show with a red dashed line, the distribution for the simulated galaxies in the test dataset. Shuffling the pixels tends to narrow the distributions around a probability value of $\sim0.5$.} \label{fig:p_shuffle}
\end{center}
\end{figure*}

The first thing to notice is that the classification applied to real data returns objects with high probability values in the 3 classes. The fraction of galaxies with all probabilities lower than 0.5 is only 2\% of the total sample. It means that the model found galaxies that resemble the galaxies in the simulation with high confidence.This reflects that the simulated galaxies are fairly similar to the observed ones and that the network found characteristic features learned in the simulations, in the CANDELS observations. Figure~\ref{fig:ex_stamps_CANDELS} shows some example stamps of observed galaxies in the three phases. It is not obvious to establish what would happen if galaxies from the training were very different from real datasets. This will be explored in future work. In order to have a first idea of how the network would behave when confronted to very different objects, we perform a simple exercise. We take the real observed galaxies from CANDELS and first randomly shuffle the central pixels of the galaxy and then shuffle all the pixels in the galaxy (inner+outskirts). This creates two fake datasets with different degrees of perturbation which are given to the network. Figure~\ref{fig:p_shuffle} shows the probability distributions for the 3 classes when these fake datasets are given. The figure shows that the first effect of shuffling the center is that the number of galaxies with a compaction probability larger than $0.5$ almost drops to zero. This is somehow expected as most of the compaction features are naturally seen in the central parts. It confirms that the network is significantly using this information to classify. Since the probabilities need to add to 1, central shuffling provokes also an increase in the number of galaxies with large probability of post-BN. Given that post-BNs tend to be more extended, the fact of shuffling the central pixels pushes the network to boost the post-BN probability since it focuses on the outer pixels. However, the values remain low ($\sim0.6$) indicating a moderate level of confidence. When both outskirts and inner pixels are shuffled, both probability distributions, BN/post-BN, significantly narrow and peak at $\sim0.4-0.5$, meaning that the network is not able to clearly assign galaxies to classes. This exercise shows that the probability distributions somehow reflect the similarity between the simulations and the observations. We notice however, that even in the shuffled images, there is a fraction of galaxies with high post-BN probabilities. A visual inspection shows that these are bright galaxies for which the shuffling has pushed bright pixels towards very large distances. The network most likely interprets this as a very extended disk. 

The fact that the distributions on CANDELS galaxies resemble to the ones obtained on the test simulated sample (red solid/dashed lines in figure~\ref{fig:p_shuffle}), suggests therefore that simulated and observed galaxies look similar to the network. This allows us to push the analysis a bit further by exploring the properties of galaxies in the three phases (BN, post-BN and pre-BN) in the observations.

\subsection{A characteristic mass range for the BN phase}
\label{sec:cmass}

\begin{figure*}
\begin{center}
$\begin{array}{c c c}
 \includegraphics[width=0.32\textwidth]{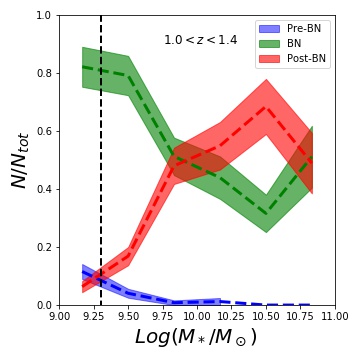} & \includegraphics[width=0.32\textwidth]{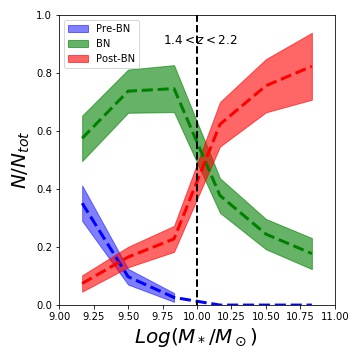} & \includegraphics[width=0.32\textwidth]{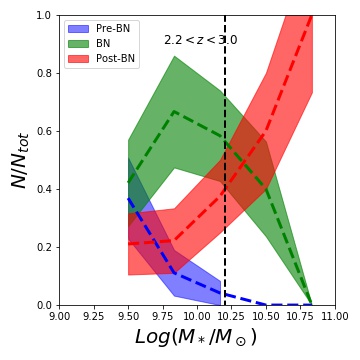}\\

\end{array}$
\caption{Stellar mass distributions of CANDELS galaxies in pre-BN (blue lines), BN (green lines) and post-BN (red lines) for different redshift bins as labelled. Galaxies in the BN phase typically peak at stellar masses of  $10^{9.2-10.3}$  as predicted by the simulations.  In more detail, the BN range is 9.5-10.3 in the high-z bin, 9.25-10.0 in
the middle-z bin, and a smaller mass in the low-z bin. This possible redshift dependence may or may not be significant. The vertical dashed lines show the mass completeness limits from Huertas-Company et al. (2016). The peak is generally below the completeness limit. This should not significantly impact the presence of the peak unless post-BN galaxies are more difficult to detect at these masses which is unlikely.} \label{fig:mass_evol}
\end{center}
\end{figure*}

In figure~\ref{fig:mass_evol} we first look at the stellar mass distributions of CANDELS galaxies in the three different phases. Recall that the simulations used for training stop at $z\sim1$, so we only show galaxies above this redshift in the observations. The abundance of galaxies in different phases strongly depends on stellar mass. Pre-BN galaxies tend to increase at low stellar masses ($M_*/M_\odot<10^{9-9.5}$) and post-BN galaxies dominate at large stellar masses ($M_*/M_\odot>10^{10.5}$). BNs are most frequent at intermediate masses and peak at $\sim10^{9.2-10.3}$. Interestingly the position of the peak seems to be relatively independent of redshift with a small tendency to move towards lower masses at lower redshifts. We notice that at this characteristic stellar mass, the CANDELS dataset is affected by incompleteness as indicated by the vertical line in the plots. This should not affect the result in the sense that there are no reasons to think that post-BN galaxies are more difficult to detect. 

This characteristic mass for compaction is a prediction from the VELA simulations as first reported in \cite{2015MNRAS.450.2327Z} and~\cite{2016MNRAS.458.4477T} and also reflected in table~\ref{tbl:table1} (see also~\citealp{2016MNRAS.457.2790T} and Dekel et al. in prep.). The fact that it appears clearly in the observations confirms that the network is automatically extracting the correlations existing in the simulations. It is worth recalling that the luminosity has been removed from the training set which ensures that the network is not classifying based on luminosity that is directly correlated with the stellar mass. The network is necessarily using other information such as spatial distribution, shape or color  to identify the different phases. The characteristic mass  naturally emerges in the observations. The network successfully identifies a population that resembles simulated galaxies experiencing compaction in the feature space learned and these galaxies tend to be near a characteristic stellar mass similar to the characteristic mass seen in the simulations.

For comparison purposes, we also show in the appendix~\ref{app:lum} the resultant mass distributions in the observations when the luminosity is left in the training set. The results are similar, confirming that luminosity is not the main parameter used by the network. There is a tendency to find more pre-BN galaxies however. We speculate that this is because the algorithm uses some S/N related information if available. Since pre-BN are generally fainter, they also have lower S/N in the observed mock images so the network will tend to classify fainter objects as pre-BN. It highlights both the strengths and risks of the deep learning approach, in the sense that all information is taken into account in our unsupervised learning. 

\begin{figure*}
\begin{center}
$\begin{array}{c c c}
 \includegraphics[width=0.32\textwidth]{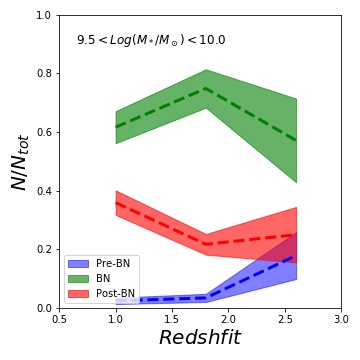} & \includegraphics[width=0.32\textwidth]{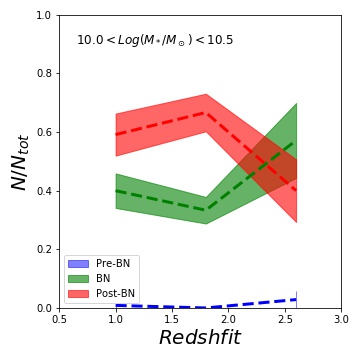} & \includegraphics[width=0.32\textwidth]{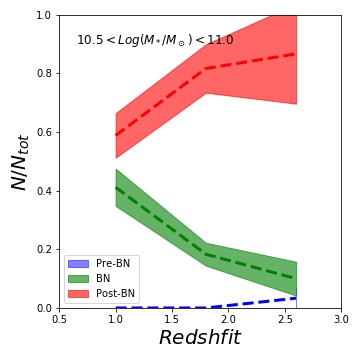}\\

\end{array}$
\caption{Redshift evolution of the fractions of CANDELS galaxies in pre-BN (blue lines), BN (green lines) and post-BN (red lines) for different stellar mass bins as labelled. In the redshift range of CANDELS ($1<z<3$), BNs dominate at a characteristic stellar mass of $\sim10^{9.2-10.4}M\odot$ as predicted by the simulations.} \label{fig:z_evol}
\end{center}
\end{figure*}

An analogous behavior is also seen in figure~\ref{fig:z_evol}, where the redshift evolution of the fractions of galaxies in the three phases at fixed stellar mass is shown. Both plots are complementary. As expected the redshift evolution strongly depends on stellar mass. The galaxies that are more frequently potentially in the BN phase in the CANDELS redshift range are in the stellar mass range of $10^{9.2}<M_*/M_\odot<10^{10.3}$. The massive compact star-forming galaxies identified in the previous works might be the high-mass tail of the BN population. More massive galaxies tend indeed to be in the post-compaction phase at all redshifts. This means that if one wants to observe the progenitors of these most massive galaxies in the process of compaction, it is required to probe $\sim10^{9.5}$ solar mass galaxies at $z>3$. That will be straightforward with JWST.

\subsection{The L-shape in sSFR vs. $M_*$}

The previous section has shown that the network  successfully identifies a characteristic galaxy stellar mass range for the BN phase in the CANDELS data. This is remarkable given the known limitations of the simulations (see section~\ref{sec:sims}) and suggests that there are important similarities between simulated and observed galaxies.

In future work, we plan to analyze in more detail how the different classes relate to other physical properties. As a preliminary step, we attempt a first look at the $sSFRs$ and central mass densities ($\Sigma_1$, \citealp{2017ApJ...840...47B}) of galaxies in pre-BN, BN and post-BN phases. This is motivated because in the simulations, the compaction, BN, and quenching sequence puts the galaxy into a characteristic  L-shape track in $sSFR-\Sigma_1$ with the BN phase at the turning point (e.g. \citealp{2015MNRAS.450.2327Z}). This L-shape is similar to the observed distribution \citep{2013ApJ...765..104B,2017ApJ...840...47B}.



\begin{figure*}
\begin{center}
$\begin{array}{c c c}
\includegraphics[width=0.33\textwidth]{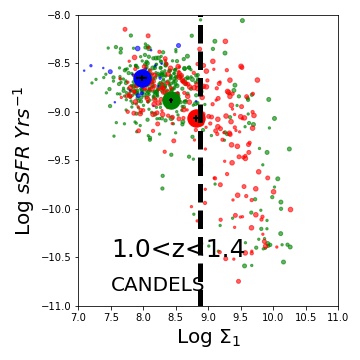} & \includegraphics[width=0.33\textwidth]{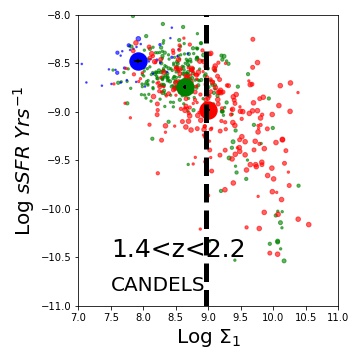}  & \includegraphics[width=0.33\textwidth]{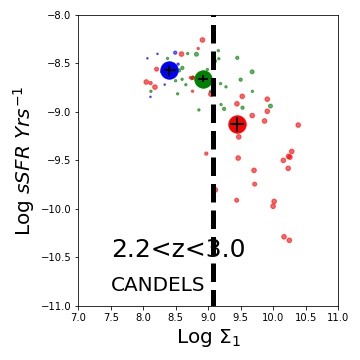} \\
\includegraphics[width=0.33\textwidth]{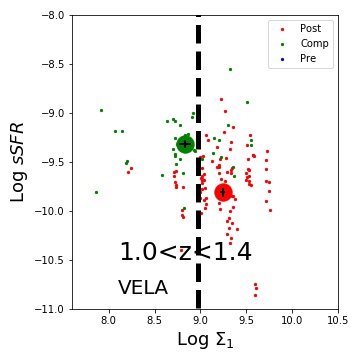} & \includegraphics[width=0.33\textwidth]{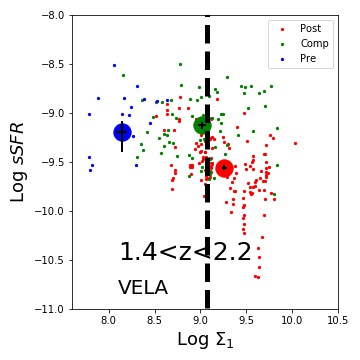}  & \includegraphics[width=0.33\textwidth]{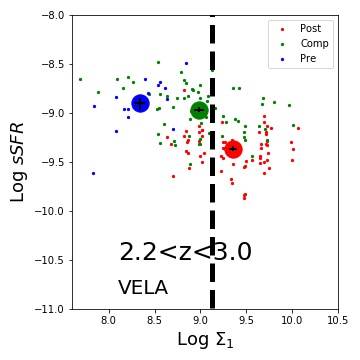} \\
\includegraphics[width=0.33\textwidth]{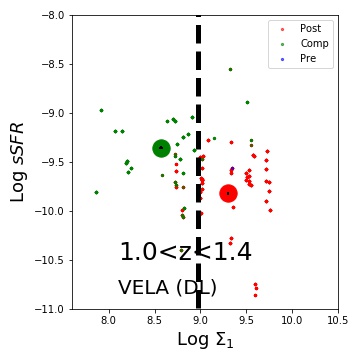} & \includegraphics[width=0.33\textwidth]{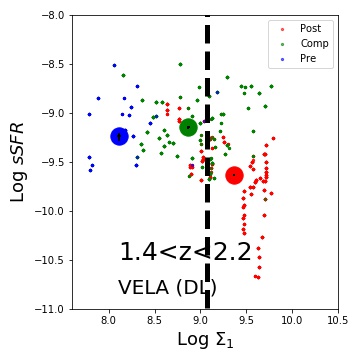}  & \includegraphics[width=0.33\textwidth]{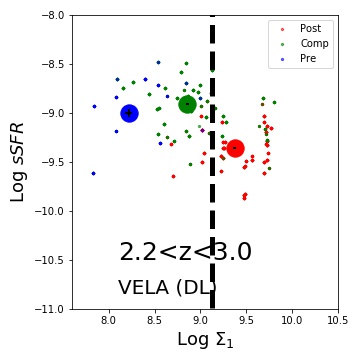} \\
\end{array}$
\caption{Distribution of pre-compaction (blue dots), compaction (green dots) and compaction (red dots) galaxies with $M_*/M_\odot>10^9$ in the $sSFR-\Sigma_1$ plane. The large dots show the average positions and the black error bars are the $68\%$ confidence interval obtained through bootstrapping. The top row show the distribution of CANDELS galaxies. The middle row are simulated galaxies with the phase defined from the assembly history. The bottom row are the same simulated galaxies when the phase is determined trough deep-learning. The vertical black dashed lines in the top row show the location of the quiescent ridge-line at a stellar mass of $10^{10}$ solar masses computed by Barro et al. 2017.  } \label{fig:delta_delta}
\end{center}
\end{figure*}

We show in figure~\ref{fig:delta_delta} the $sSFR-\Sigma_1$ plane for pre-BN, BN and post-BN galaxies in CANDELS as defined by the CNN trained on the simulations. As previously reported, galaxies form a characteristic L-shape distribution in the plane. 


At first approximation, the median position (large dots in the figure) of pre-BN, BN and post-BN galaxies is different, and crudely follow the expected evolutionary sequence from the simulations. Pre-BN galaxies tend to be indeed in the main-sequence and have low central density values while post-BN galaxies have lower specific star-formation rates and larger central densities. BN galaxies lie in between. Given the observability  timescales calibrated in section~\ref{sec:obs_time}, this suggests that there is an evolutionary sequence in the plane and that galaxies tend to move from left to right.  We observe however that there is also significant overlap between the different phases in the three quadrants of the $sSFR-\Sigma_1$ diagram. For example, several galaxies are classified as post-BN while they have low $\Sigma_1$ values. Also, there is mixing of low sSFR and high sSFR compact galaxies that is not fully consistent with the distinction between the BN and post-BN phases in the simulations. For comparison, we show the same plot for the VELA simulations which shows a clearer separation, namely a stronger correlation between the three phases as defined based on the gas/SFR distribution and the distribution to three quadrants in the $sSFR-\Sigma_1$ diagram as derived from the stellar distribution.

We emphasize that the main purpose of this work is to illustrate the methodology.  We thus keep for future work a detailed investigation of the reasons of this increased confusion in CANDELS. One possible explanation resides in the definition of the BN phase used for training. We recall that several galaxies in the simulation present complex assembly histories, with many wet-compaction events of different intensities (see figure~\ref{fig:comp_def}). A similar behavior is also reported in \cite{2016MNRAS.457.2790T}, i.e. compaction and quenching events confine the galaxy to the main sequence, until a major BN event that is followed by long-term quenching as a result of a hot massive halo. Therefore, according to our labelling of the training set explained in section~\ref{sec:labels}, galaxies can be still considered as post-BN (see for example VELA11 in figure~\ref{fig:pred_time}) in between several events which could also contribute explaining the overlap we see in CANDELS. A way to explore the effects of minor compaction events would be to train a network with only major compactions and see how the classification changes. To do that a larger and more diverse training set is needed and also at higher redshift, in the JWST range, where major events tend to happen in the simulations. We keep this for future work.

\section{Summary and conclusions}

We have explored a new approach to classify galaxy images using deep learning calibrated on numerical simulations.  The general methodology consists first of generating mock images of galaxies reproducing the observing conditions from hydro cosmological simulations which are then labelled based on the known evolution of gas, SFR and stars. The images are then fed to an unsupervised feature learning machine that automatically  learns the features to detect a given evolution pattern. We have applied the method for detecting the characteristic blue nugget (BN) phase as seen in cosmological simulations, near a critical mass and preferentially at high redshifts, following a wet compaction process and followed by central quenching. We have used for that purpose a suite of high resolution zoom-in hydro numerical simulations of intermediate mass galaxies in the redshift range $1<z<3$. We have shown that a simple CNN is able to detect galaxies in the BN phase with $\sim80\%$ accuracy within a time window of $\pm0.2$ Hubble times and hence establish temporal constraints in the data. The described methodology presents several key advantages over more traditional approaches. First of all, it does not require any image pre-processing. Only the pixel distributions are fed into the network which automatically extracts the relevant information. This does not prevent however to combine the automatically extracted features with other standard measurements such as colors or sizes. Moreover there is no need of an a-priori assumption of the \emph{optimal} observables for a given physical process. The procedure will automatically extract the best tracers if present in the data.\\

We have then applied the trained model to observed galaxy multicolor images from the CANDELS survey observed with HST in the same redshift range and classify them into three main classes: pre-BN, BN and post-BN. 

The key results are:
\begin{itemize}
\item The network finds galaxies with high probability of being in the three classes indicating  similarity between simulated and observed galaxies. 
\item The classification recovers a characteristic stellar mass for the BN phase of $\sim10^{9.2-10.3}$ solar masses mostly independent of redshift.  More massive compact galaxies are found to be preferentially in the post-BN class, so they are compatible with having gone through the BN phase more than $0.5$ Hubble times before the time of observation. 
\item Pre-BN, BN and post-BN galaxies occupy different regions in the $sSFR-\Sigma_1$ plane, suggesting an evolutionary sequence in the plane as predicted by the simulations. There is however some degree of confusion, i.e. post-BN galaxies with low central densities that will be investigated in future work. 

\end{itemize}

 In particular, one important point that will be addressed in forthcoming works is the impact of the specific set of simulations used for training. Despite the similarities between simulations and observations suggested in section~\ref{sec:cmass}, the VELA simulations used in this work might be still too limited to adequately represent the entire CANDELS data set, not only because of the lack of AGN but also because the sample is small and covers a limited mass range. Additionally, the assumptions regarding the sub-grid astrophysics are not well constrained by theory or observations as discussed in section~\ref{sec:sims}. To further investigate the impact of these limitations, we plan to enlarge our training sets by using new available simulated datasets with the same VELA initial conditions but different sub-grid astrophysics as well as other independent simulated datasets including AGN.  

The presented methodology could then be adapted to other robust physical processes captured in simulations and could constitute a useful tool to better compare future imaging surveys with forthcoming simulations.


\section*{Acknowledgments}

The authors are grateful to Google for the unrestricted gift given to the University of California Santa Cruz to carry out the project: "deep learning for Galaxies" that greatly contributed to make this work possible. We also appreciate helpful discussions with Sander Dielemann, Daniel Freedman, Eric Hayashi and Jon Shlens at Google. We also thank Fr\'ed\'eric Bournaud for refereeing this work and providing interesting suggestions. This work was partly supported by the grants France-Israel PICS, US-Israel BSF 2014-273, and NSF AST-1405962. JRP acknowledges support from HST-AR-14578.001-A. AD also acknowledges support from GIF I-1341-303.7/2016, DIP STE1869/2-1 GE625/17-1, and I-CORE PBC/ISF 1829/12. MHC acknowledges support from the ANR ASTROBRAIN.  DC has been funded by the ERC Advanced Grant, STARLIGHT: Formation of the First Stars (project number 339177). The VELA simulations were performed at the National Energy Research Scientific Computing Center (NERSC) at Lawrence Berkeley National Laboratory, and at NASA Advanced Supercomputing (NAS) at NASA Ames Research Center.

\appendix

\section{The effect of luminosity}
\label{app:lum}

\begin{figure*}[h]
\begin{center}
$\begin{array}{c c c}
 \includegraphics[width=0.32\textwidth]{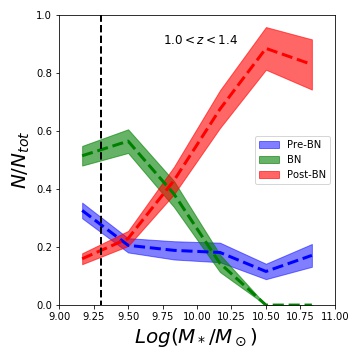} & \includegraphics[width=0.32\textwidth]{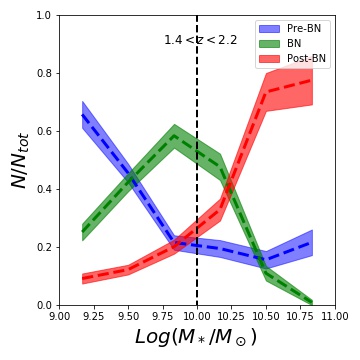} & \includegraphics[width=0.32\textwidth]{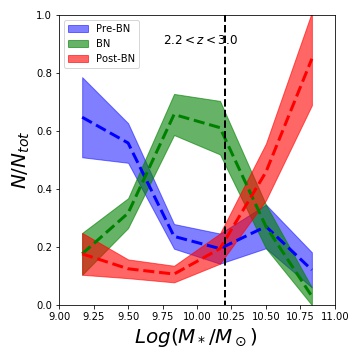}\

\end{array}$
\caption{Stellar mass distributions of CANDELS galaxies in pre-BN (blue lines), BN (green lines) and post-BN (red lines) for different redshift bins as labelled. The classification is performed with a training set including the luminosity information (see text for details). Galaxies in the BN phase typically peak at stellar masses of  $10^{9.2-10.3}$ at all redshifts.  The vertical dashed lines show the completeness limits from Huertas-Company et al. (2016). } \label{fig:mass_evol_lum}
\end{center}
\end{figure*}

In the training set used in this work, the magnitudes of the galaxies in the different phases were randomly changed. This is to ensure that all galaxies have similar S/N and that the network does not learn based on that. As a matter of fact, since pre-BN galaxies  in the simulations are found at higher redshift and have lower stellar masses than post-BN galaxies, they will be more noisy in the \emph{CANDELized} images. The network might therefore use this information. To check the effect of this in the final classification, we show in figure~\ref{fig:mass_evol_lum} the same stellar mass distributions of galaxies in the three different phases in CANDELS as in figure~\ref{fig:mass_evol} but obtained with a training set without randomizing the magnitudes. As can be seen, the distribution is similar, i.e. a BN peak at a characteristic stellar mass. However, the code tends to find more pre-BN galaxies at low mass. This is because it is learning some information from the S/N distribution. This exercice shows the strength of the deep-learning approach since it demonstrates that the network uses all available information. It highlights however the risks too. One needs to control the information that should not be used by the net. 

\end{document}